\documentclass [twocolumn, aps, showpacs] {revtex4}
\usepackage{graphicx}
\begin{document}
\draft
\title {Role of finite layer thickness in spin-polarization of GaAs 2D electrons in strong parallel magnetic fields}

\author {E. Tutuc, S. Melinte, E.P. De Poortere, M. Shayegan}
\address{Department of Electrical Engineering, Princeton
University, Princeton, NJ 08544}
\author {R. Winkler}
\address{ Institut f\"ur Technische Physik III, Universit\"at
Erlangen-N\"urnberg, Staudtstr. 7, D-91058 Erlangen, Germany}
\date{\today}
\begin{abstract}
We report measurements and calculations of the spin-polarization,
induced by a parallel magnetic field, of interacting, dilute,
two-dimensional electron systems confined to GaAs/AlGaAs
heterostructures. The results reveal the crucial role the non-zero
electron layer thickness plays: it causes a deformation of the
energy surface in the presence of a parallel field, leading to
enhanced values for the effective mass and g-factor and a
non-linear spin-polarization with field.
\end{abstract}
 \maketitle

The spin-polarization of an interacting, dilute two-dimensional
(2D) carrier system has been of interest for decades.  It has long
been expected that because of Coulomb interaction the product
$g^{\ast}m^{\ast}$, which determines the spin susceptibility of
the 2D system, increases as the 2D density ($n$) is lowered and
eventually diverges as the system makes a transition to a
ferromagnetic state at sufficiently low $n$
\cite{tanatar,attaccalite} ($g^{\ast}$ and $m^{\ast}$ are the
carrier Land\'e g-factor and effective mass, respectively).
Recently, there has been much renewed interest in this problem,
thanks to the availability of high-quality dilute 2D systems, and
the belief that it may shed light on the controversial issue of a
metal-insulator transition in 2D \cite{abrahams}. A technique
commonly used to study the spin-polarization is to measure the
response of the 2D system to a tilted or parallel magnetic field
\cite{okamoto,tutuc,vitkalov,pudalov,tutuc02,proskuryakov,noh,etienne}.
The results of some of these measurements
\cite{tutuc02,proskuryakov,noh}, however, appear to be at odds
with what is theoretically expected \cite{tanatar,attaccalite} for
a dilute, interacting 2D system that is otherwise {\it ideal},
i.e., has zero layer thickness and is disorder-free. In
particular, when $g^{\ast}m^{\ast}$ is deduced from parallel
magnetic field at which the 2D system becomes fully
spin-polarized, then the experimental results for GaAs 2D
electrons \cite{tutuc02} and holes \cite{proskuryakov,noh} suggest
a decreasing value of $g^{\ast}m^{\ast}$ with decreasing $n$,
opposite to the theoretical predictions.

Here we report a combination of measurements and calculations for
the parallel magnetic field-induced spin-polarization of 2D
electrons at the GaAs/AlGaAs heterojunction. The results highlight
the importance of the finite thickness of the electron layer and
the resulting deformation of the energy surface
$E(\mathbf{k}_\|)$, where $\mathbf{k}_\|$ is the in-plane wave
vector, that occurs in the presence of a strong parallel field.
This deformation induces an enhancement of both $m^{\ast}$ and
$g^{\ast}$ and leads to a {\it non-linear} spin-polarization in a
parallel field. We find that, once the effect of the finite layer
thickness and interaction is taken into account, there is
reasonable agreement between the experimental data and
calculations.

We used five samples from three different wafers (A, B,
and C). The samples were all modulation-doped GaAs/AlGaAs
heterojunctions with $n$ in the range $0.8$ to $6.5\times10^{10}$
cm$^{-2}$. Their low-temperature mobility varied
depending on the sample and $n$; at $n=2\times10$ cm$^{-2}$, it ranged from about $2\times10^{5}$ to $2\times10^{6}$
cm$^{2}$/Vs. Samples were patterned in
either van der Pauw or Hall bar shapes, and were fitted with back-
or front-gates. To tune $n$ in samples A, B1, and B2, following
illumination with a red LED, we used front-gate bias; for B3 (in
the range $n < 4\times10^{10}$ cm$^{-2}$) and C we used back-gate
bias and no illumination. For B3, the highest density ($n =
6.5\times10^{10}$ cm$^{-2}$) was obtained after illumination,
followed by back-gating to reduce $n$ to $4.5\times10^{10}$
cm$^{-2}$. Measurements were done down to a temperature of 30 mK,
and a rotating platform was used to tune the angle between the
applied magnetic field and the sample plane.

Figure 1 summarizes our data taken on the different samples.
Plotted are the values of $g^{\ast}m^{\ast}/g_{b}m_{b}$,
determined from the parallel magnetic field, $B_{P}$, at which the
2D system becomes fully spin polarized ($m_{b}=0.067m_{0}$ and
$g_{b}=-0.44$ are the band effective mass and Land\'{e} g-factor
for GaAs electrons; $m_{0}$ is the free electron mass). The
parallel magnetic field ($B_{\|}$) leads to the formation of two
energy subbands, one for each spin, and separated by the Zeeman
energy, $E_{Z}=|g^{\ast}|\mu_{B}B_{\|}$, where $\mu_{B}$ is the
Bohr magneton. The 2D system becomes fully spin-polarized above a
field $B_{P}$ at which $E_{Z}$ equals the Fermi energy. The
equality leads to an expression for $B_{P}$: $B_{P} =
(h^{2}/2\pi\mu_{B}) \cdot (n/|g^{\ast}|m^{\ast})$, from which we
determine $g^{\ast}m^{\ast}$ that are plotted in Fig.\ 1.

\begin{figure}
\centering
\includegraphics[scale=0.33]{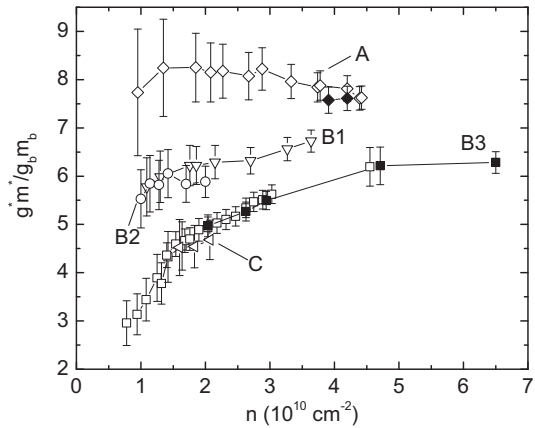}
\caption{Values of $g^{*}m^{*}/g_{b}m_{b}$, determined from the
parallel magnetic field $B_{P}$ at which the 2D electrons in GaAs
become fully spin-polarized, are shown as a function of 2D
electron density. Results are shown for five samples with
low-temperature mobilities (at $n=2\times10^{10}$ cm$^{-2}$) of A:
4, B1: 16, B2: 16, B3: 7, C: $2\times10^{5}$ cm$^{2}$/Vs. The
closed symbols represent $B_{P}$ determined from SdH measurements
in a nearly parallel magnetic field while open symbols are for
$B_{P}$ from magnetoresistance measurements in a parallel field.}
\end{figure}

The procedures we have used to experimentally determine $B_{P}$
have been described elsewhere \cite{tutuc,tutuc02}; here we give a
brief summary. We determine $B_{P}$ from two independent sets of
experiments: Shubnikov - de Haas (SdH) measurements in a nearly
parallel magnetic field, and magnetoresistance measurements in a
strictly parallel field. In the first type of experiment, we apply
a constant magnetic field ($B_{tot}$) whose initial direction is
parallel to the 2D electron plane, and then slowly rotate the
sample while recording the sample resistance as a function of the
angle between the 2D plane and the field direction. If we limit
ourselves to small angles, the field's parallel component
($B_{\|}$) remains almost constant (equal to $B_{tot}$ typically
to better than 1$\%$) during the rotation, while its perpendicular
component ($B_{\bot}$) changes sufficiently to probe the SdH
oscillations. We then Fourier analyze the SdH oscillations to
obtain the populations of the two spin subbands. These densities
provide a direct measure of the spin-polarization of 2D electron
system and allow us to determine the field $B_{P}$ above which the
system becomes fully spin-polarized. In the second type of experiment
we measure the sample
resistance as a function of a magnetic field applied strictly in
the 2D plane. As shown elsewhere \cite{tutuc,tutuc02}, the
in-plane magnetoresistance shows a marked change in
its functional form at the field $B_{P}$. For a given sample, the
value of $B_{P}$ obtained from the two types of experiments (open
and closed symbols in Fig.\ 1) are in agreement.

Data of Fig.\ 1 illustrate that the product $g^{\ast}m^{\ast}$,
deduced from $B_{P}$ as described in the last two paragraphs,
deviates substantially from what is expected for an ideal,
interacting 2D electron system, namely a monotonically increasing
$g^{\ast}m^{\ast}$ as $n$ is lowered \cite{tanatar,attaccalite}.
The data also reveal that the measured $g^{\ast}m^{\ast}$ is
sample dependent and not a unique function of $n$. A possible
reason for this non-uniqueness may be the sample disorder that
indeed varies between different samples. An examination of the
data, however, argues against this hypothesis: considering the
data at a given density, it is clear that there is no simple trend
linking the sample disorder, as deduced from the low-temperature
mobility, to the measured $g^{\ast}m^{\ast}$. As we demonstrate
below, another factor that renders the experimental 2D electrons
non-ideal, namely their finite layer thickness, appears to be
responsible for the sample-dependent $g^{\ast}m^{\ast}$ and the
difference between the observed and expected density dependence of
$g^{\ast}m^{\ast}$.

In the presence of a large $B_{\|}$, when the magnetic length ($=
\sqrt{\hbar/eB_{\|}}$) becomes comparable to or smaller than the
thickness of the electron layer, the energy surface
$E(\mathbf{k}_\|)$ of the electrons gets deformed in the in-plane
direction perpendicular to $B_\|$.  The deformation leads to an
increase of the in-plane effective mass, $m^{\ast}$, which, in
second order perturbation theory, is given by \cite{stern}
\begin{equation}
  \label{eq:mstar}
  m^\ast (B_\|) = m_{b} \bigg/ \sqrt{1 - \frac{2e^2 B_\|^2}{m_{b}}
  \sum_{j\ne 0} \frac{|\langle z \rangle_{0j} |}{E_j - E_0}}
\end{equation}
where the sum runs over all excited subbands $j$ and $z$ is the
quantization axis  \cite{das sarma}. Data of Fig.\ 2 provide an
experimental demonstration of this effect in our samples. In Fig.\
2(a) we show, as a function of $B_{\|}$, the measured
magnetoresistance of samples A and B1, both at a density of
$2.7\times10^{10}$ cm$^{-2}$. The magnetoresistance for each
sample shows a clear change in its dependence on $B_{\|}$ at a
field marked by a vertical arrow as $B_{P}$. As demonstrated
previously \cite{tutuc,tutuc02}, the field $B_{P}$ marks the onset
of full spin-polarization. Note in Fig.\ 2(a) that $B_{P}$ is
larger for sample B1 than A even though they have the same
density. We also measured $m^{\ast}$ in the two samples as a
function of $B_{\|}$, as shown in Fig.\ 2(b) \cite{footnote1}. We
determined $m^{\ast}$ from the temperature dependence of the
amplitude of the SdH oscillations, measured as the sample was
slowly rotated in an almost parallel field. We performed a
standard analysis, fitting the amplitude of the SdH oscillations
($\Delta R$) to the Dingle formula, $\Delta R \sim \xi/\sinh\xi$,
where $\xi\equiv2\pi^{2}k_{B}T/\hbar\omega_{c}$ and
$\omega_{c}=eB/m^{\ast}$. It is clear in Fig.\ 2(b) that
$m^{\ast}$ for both samples exhibits a strong enhancement with
increasing $B_{\|}$, consistent with Eq.\ (\ref{eq:mstar}). Moreover, $m^{\ast}$ has a larger enhancement for sample A
than for B1. This stronger enhancement correlates with the smaller
$B_{P}$ measured for sample A. We believe that the main difference
between the two samples in Figs.\ 2(a) and (b) is that sample A
has a larger layer thickness: because of a decrease in subband
separation, the larger thickness leads to the larger enhancement
of $m^{\ast}$, consistent with the smaller $B_{P}$ for sample A
compared to B1.

\begin{figure}
\centering
\includegraphics[scale=0.66]{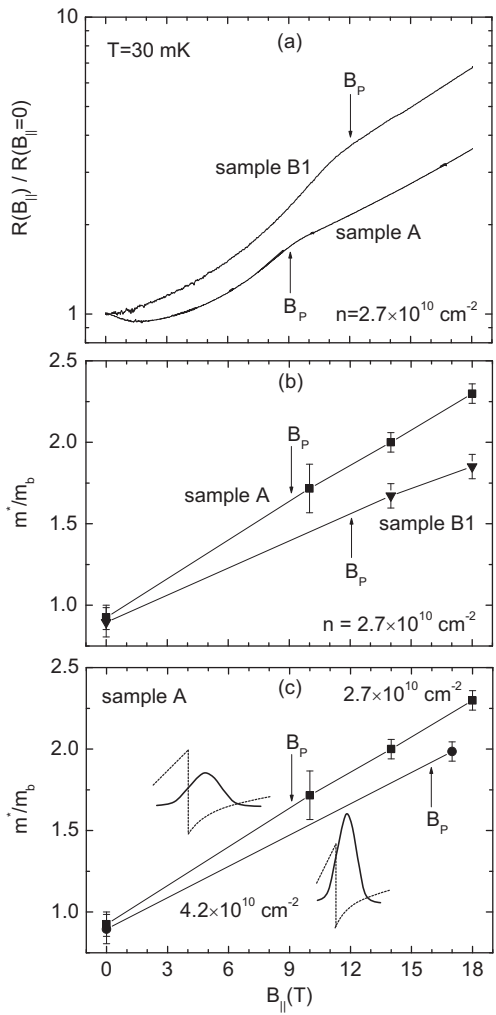}
\caption{(a) Parallel-field magnetoresistance of samples A and B1
at $2.7\times10^{10}$ cm$^{-2}$. (b) Effective masses measured for
the two samples of panel (a). (c) Effective masses measured for
sample A at two different densities. The density is tuned via a
front-gate bias which leads to a narrowing of the wavefunction at
higher density, as shown schematically in the insets. In all three
panels, the fields $B_{P}$ above which the 2D electrons become
fully spin-polarized are marked by vertical arrows.}
\end{figure}

To further substantiate the connection between layer thickness and
$m^{\ast}$ enhancement, in Fig.\ 2(c) we show data for sample A at
a higher carrier density of $4.2\times10^{10}$ cm$^{-2}$. The
measured $m^{\ast}$ enhancement is smaller for the higher $n$
state. This is consistent with layer thickness being responsible
for the $m^{\ast}$ enhancement: as we use a more positive front-gate
bias to increase $n$ in this sample, the electron
wavefunction is squeezed more towards the interface so that the
layer thickness is reduced [see the insets to Fig.\ 2(c)],
consistent with the smaller measured $m^{\ast}$ enhancement.

\begin{figure}
\centering
\includegraphics[scale=0.66]{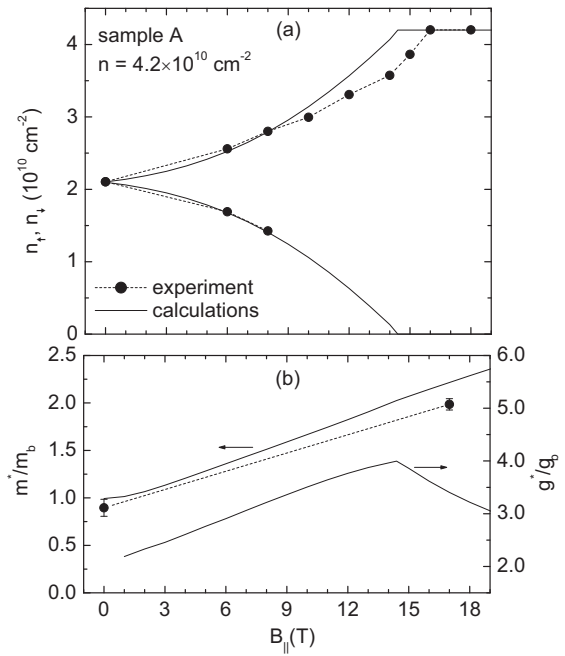}
\caption{Results of calculations are shown for (a) spin subband
densities $n_{\uparrow}$ and $n_{\downarrow}$, (b) effective mass,
and the g-factor for a GaAs 2D electron system (density
$4.2\times10^{10}$ cm$^{-2}$) with realistic finite layer
thickness. Also plotted (circles) are the experimentally measured
$m^{\ast}$, $n_{\uparrow}$ and $n_{\downarrow}$ at this density.}
\end{figure}

In order to quantitatively understand the experimental data, we
have done self-consistent density-functional calculations of the
subband structure in the presence of $B_\|$. We used the recent
parameterization of the exchange-correlation energy by Attaccalite
{\em et al.}\ \cite{attaccalite}. The energy and length scales for
electrons in a semiconductor are characterized by the effective
Rydberg and the Bohr radius according to the effective mass
$m^\ast$ and the dielectric constant of the material.  In our
calculations it was crucial that $m^\ast$ was determined as a
function of $B_\|$ from the self-consistently calculated subband
dispersion $E(\mathbf{k}_\|, B_\|)$. These calculations confirm
the qualitative trends expected from Eq.\ (\ref{eq:mstar}). Band
structure effects beyond the effective-mass approximation are not
consisdered here. We have checked that these effects are of minor
importance \cite{winkler93}. In the calculations, the field $B_P$
is defined as the smallest value of $B_{\|}$ for which the fully
spin-polarized configuration has the lowest total energy.

Figure 3 provides an example of the results of the calculations
for $n=4.2\times10^{10}$ cm$^{-2}$; shown are (a) the spin-subband
densities $n_{\uparrow}$ and $n_{\downarrow}$, (b) $m^{\ast}$, and
(c) $g^{\ast}$, as a function of $B_{\|}$. The calculations were
done using the parameters of sample A (spacer thickness and
barrier height), assuming a p-type background doping of
$2.7\times10^{13}$ cm$^{-3}$, and a binding energy of 90 meV for
the dopants (Si) in the barrier. These are reasonable values,
consistent with our estimate of the unintentional (residual)
doping in our molecular beam epitaxy system and the binding
energies quoted in the literature \cite{schubert}. The
calculations predict a {\it non-linear} but smooth increase of the
spin-polarization as a function of $B_{\|}$.

In Figs.~3 (a) and (b) we have also included our measured
spin-subband densities and $m^{\ast}$, determined from SdH
oscillations in a nearly parallel magnetic field. Overall, there
is good agreement between the experimental data and calculations
for both spin-polarization and $m^{\ast}$ as a function of
$B_{\|}$. The calculated $B_{P}=14.5$~T agrees well with the
measured $B_{P}\cong 16$~T, and is much smaller than $B_{P}\cong
48$~T expected for an interacting GaAs 2D electron system with
zero layer thickness at $n=4.2\times 10^{10}$ cm$^{-2}$ (see
dashed curve in Fig. 4).

\begin{figure}
\centering
\includegraphics[scale=0.33]{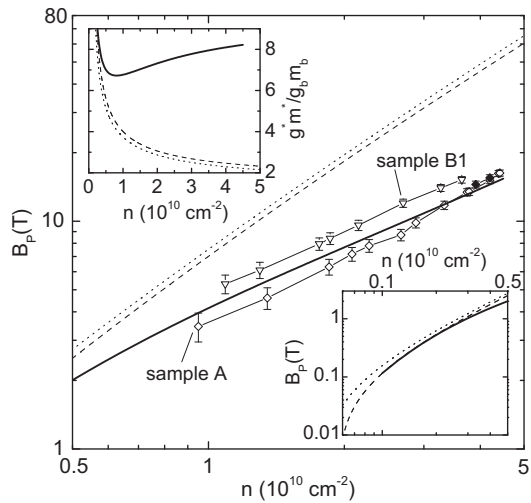}
\caption{Comparison between the measured $B_{P}$ for samples A and
B1 (same symbols as in Fig.\ 1) and calculations that take the
finite layer thickness of the GaAs 2D electrons into account
(solid curve). The density was varied via a front-gate bias in
both the experiments and calculations. For contrast, we also show
$B_{P}$ from two sets of calculations that are based on Ref.
\cite{attaccalite} and are done for interacting, zero layer
thickness 2D electrons in GaAs (see text). The lower inset shows
the calculated curves of the main panel at lower densities. The
upper inset shows $g^{*}m^{*}$ based on different calculations.}
\end{figure}

To understand the density dependence of $B_{P}$ for a
given sample, we also calculated $B_{P}$ as a function of $n$,
which is tuned either with a front- or back-gate bias. For these
calculations, we kept the sample parameters fixed, and only
changed the boundary conditions for the Hartree potential thus
simulating the effect of the gate bias. The results for the
case where the front-gate is used are shown in Fig.\ 4 (solid
curve), along with the experimental data for samples A and B1.
There is reasonable agreement between the calculations and the
experimental data \cite{footnote2}.

In Fig.\ 4 we also plot $B_{P}$ vs. $n$ (dashed curve) calculated
based on Ref.\ \cite{attaccalite} where an interacting 2D electron
system with zero layer thickness is assumed, as well as $"B_{P}"$
(dotted curve) determined from the calculated spin susceptibility
in the limit $B_{\|}=0$ \cite{attaccalite} and {\it assuming} a
{\it linear} spin-polarization as a function $B_{\|}$. Three
noteworthy trends are observed in Fig. 4. First, for
$n>0.5\times10^{10}$ cm$^{-2}$ the dashed and dotted curves are
very close to each other, meaning that the spin polarization of an
ideal, zero-thickness 2D system is approximately (within 5\%)
linear with $B_{\|}$ in this density range or, equivalently, the
Zeeman splitting can be expressed well in terms of an effective
g-factor independent of $B_{\|}$. This implies that the finite
layer thickness is the key factor that leads to the observed
non-linearity in spin-polarization with $B_{\|}$ and the resulting
reduction of $B_{P}$. The mechanism responsible for this
non-linearity can be summarized as follows: $B_{\|}$ induces an
increase of $m^{\ast}$ due to the finite layer thickness. This has
a twofold effect. It directly reduces $B_P$ because $B_P \propto
1/m^{\ast}$. Furthermore, the increase of $m^{\ast}$ reduces the
effective Bohr radius and thus increases $r_{s}$, the average
electron spacing measured in units of effective Bohr radius. The
increase in $r_{s}$ in turn yields an increase of $g^{\ast}$ due
to the Coulomb interaction. Second, the solid and dashed curves
merge as the density is lowered, consistent with the expectation
that, because of the smaller $B_{P}$, the finite layer thickness
induced $g^{\ast}m^{\ast}$ enhancement becomes less important.
Third, at ultra-low densities the dashed and dotted curves start
to diverge (see the lower inset), implying that the interaction
alone can induce a significant non-linearity of spin-polarization
with $B_{\|}$ in a very dilute 2D system \cite{attaccalite}.

Finally, if we use the expression
$B_{P}=(h^{2}/2\pi\mu_{B})\cdot(n/|g^{\ast}|m^{\ast})$ to
determine $g^{\ast}m^{\ast}$ from $B_{P}$ as frequently done in
the literature, we obtain the curves shown in Fig. 4 upper inset.
These plots emphasize that "$g^{\ast}m^{\ast}$" deduced from
$B_{P}$ for a 2D system with finite layer thickness (solid curve)
is significantly enhanced with respect to the ideal 2D system and
can show a non-monotonic dependence on $n$. The results therefore
caution against extracting values for $g^{\ast}m^{\ast}$ {\it in
the limit of zero magnetic field} from measurements of $B_{P}$ at
large parallel fields \cite{zhu}. Moreover, $B_{P}$ and
$g^{\ast}m^{\ast}$ are not unique functions of $n$; they depend on
the electron layer thickness which in turn depends on sample
parameters and experimental conditions.

We thank NSF and DOE for support, D.M. Ceperley and S. Moroni for
helpful discussions. Part of this work was done at NHMFL; we thank
T. Murphy and E. Palm for support.


\begin{thebibliography}{10}
\small

\bibitem{tanatar}
B. Tanatar and D.M. Ceperley, \prb {\bf 39}, 5005 (1989), and
reference therein.

\bibitem{attaccalite}
C. Attaccalite {\it et~al.}, \prl {\bf 88}, 256601 (2002).

\bibitem{abrahams}
E. Abrahams {\it et~al.}, Rev. Mod. Phys. {\bf 73}, 251(2001).

\bibitem{okamoto}
T. Okamoto {\it et~al.}, \prl {\bf 82}, 3875 (1999).

\bibitem{tutuc}
E. Tutuc {\it et~al.}, \prl {\bf 86}, 2858 (2001).

\bibitem{vitkalov}
S.A. Vitkalov {\it et~al.}, \prl {\bf 87}, 086401 (2001); A.A.
Shashkin {\it et al.}, \prl {\bf 87}, 086801 (2001).

\bibitem{pudalov}
V.M. Pudalov {\it et~al.}, \prl {\bf 88}, 196404 (2002).

\bibitem{tutuc02}
E. Tutuc {\it et~al.}, \prl {\bf 88}, 036805 (2002).

\bibitem{proskuryakov}
Y.Y. Proskuryakov {\it et~al.}, \prl {\bf 89}, 076406 (2002).

\bibitem{noh}
Hwayong Noh {\it et~al.}, cond-mat/0206519.

\bibitem{etienne}
E.P. De Poortere {\it et~al.}, \prb {\bf 66}, 161308 (2002).

\bibitem{stern}
F. Stern, \prl {\bf 21}, 1687 (1968).

\bibitem{das sarma}
The role of finite layer thickness in determining the
$B_{\|}$-dependence of the magnetoresistance of 2D carrier systems
was recently reported [S. Das Sarma and E.H. Hwang, \prl {\bf 84},
5596 (2000)].

\bibitem{footnote1}
In our $m^{\ast}$ measurements, we chose $B_{\|}>B_{P}$ so that
only one spin subband is occupied. In this case the Landau levels
are simply separated by $\hbar\omega_{c}\equiv\hbar
eB_{\bot}/m^{*}$.

\bibitem{winkler93}
R. Winkler and U.\ R\"ossler, \prb {\bf 48}, 8918 (1993).

\bibitem{schubert}
E.  Schubert and K. Ploog, \prb {\bf 30}, 7021 (1984).

\bibitem{footnote2}
In both experimental data (Fig.\ 1) and calculations (not shown),
$B_{P}$ decreases less quickly when a back-gate (sample B3),
rather than a front-gate (samples A and B1), is used to decrease
$n$. This is consistent with the finite layer thickness effect.
When we use a front-gate to reduce the density, the wavefunction
gets thicker while the opposite is true when the back-gate is
used.

\bibitem{zhu}
Also see J. Zhu {\it et~al.}, cond-mat/0301165.

\end{thebibliography}
\end{document}